\documentclass[review]{elsarticle}

\usepackage{lineno,hyperref}
\usepackage{graphicx}
\usepackage{subfig}
\usepackage[linesnumbered,ruled,lined]{algorithm2e}
\usepackage{float}
\usepackage{tikz}
\usepackage{epstopdf}
\usepackage{amssymb}
\usepackage{amsmath}
\usepackage{booktabs}
\usepackage{array}
\usepackage{multirow}
\usepackage{mathtools}
\usetikzlibrary{arrows.meta}
\usepackage{natbib}
\usepackage{upgreek}
\usepackage{textcomp}

\journal{X. X. X}

\begin{document}

\begin{frontmatter}
	
	\title{Generative Adversarial Networks with Physical Evaluators for Spray Simulation of Pintle Injector}
	\author[Hao's address]{Hao Ma}
	\ead{hao.ma@tum.de}		
	\author[Botao's address]{Botao Zhang}
	\ead{2018zhangbotao@mail.nwpu.edu.cn}
	\author[Chi's address]{Chi Zhang}
	\ead{c.zhang@tum.de}
	\author[Hao's address]{Oskar J. Haidn \corref{mycorrespondingauthor}}
	\ead{oskar.haidn@tum.de}
	\cortext[mycorrespondingauthor]{Corresponding author. Tel.:.+49 89 289 16084.}
	
	\address[Hao's address]{Department of Aerospace and Geodesy, Technical University of Munich, 85748 Garching, Germany}
	\address[Botao's address]{Key Laboratory for Liquid Rocket Engine Technology, Xi'an Aerospace Propulsion Institute, 710100 Xi'an, China}
	\address[Chi's address]{Department of Mechanical Engineering, Technical University of Munich, 85748 Garching, Germany}

\begin{abstract}
Due to the adjustable geometry, pintle injectors are specially suitable for the liquid rocket engines which require a widely throttleable range.
While applying the conventional computational fluid dynamics approaches to simulate the complex spray phenomena in the whole range still remains to be a great challenge.
In this paper, a novel deep learning approach used to simulate instantaneous spray fields under continuous operating conditions is explored.
Based on one specific type of neural networks and the idea of physics constraint, a \emph{Generative Adversarial Networks with Physics Evaluators} (GAN-PE) framework is proposed.
The geometry design and mass flux information are embedded as the inputs. 
After the adversarial training between the generator and discriminator, the generated field solutions are fed into the two physics evaluators.
In this framework, mass conversation evaluator is designed to improve the training robustness and convergence. 
And the spray angle evaluator, which is composed of a down-sampling CNN and theoretical model, guides the networks generating the spray solutions more according with the injection conditions.
The characterization of the simulated spray, including the spray morphology, droplet distribution and spray angle, is well predicted.
The work suggests a great potential of the prior physics knowledge employment in the simulation of instantaneous flow fields.

\end{abstract}


\end{frontmatter}

\clearpage

\section{Introduction}\label{sec:Introduction}

Due to a wider throttling range and greater combustion stability, pintle injectors are specially suitable for the liquid rocket engines that require deep, fast, and safe throttling\cite{dressler2000trw,heister2011pintle}, such as the descent propulsion system in Apollo program\cite{hammock1973apollo} and the reusable Merlin engine of SpaceX\cite{bjelde2008spacex}.

In the practical throttleable engine applications, the pintle is movable to alter the injection area so that the mass flow rate of the injected propellants can be varied continuously according to the economical and safe thrust curve in a given situation\cite{casiano2010liquid}.
However, in the previous spray simulations of pintle injectors, the changes were only considered under discrete condition combinations over a limited amount of select operating points\cite{radhakrishnan2018lagrangian,son2016verification,son2017numerical}.
For the traditional discrete methods they used, simulations have to be conducted repeatedly to varies the operating conditions and the computational cost becomes prohibitively expensive\cite{yu2019flowfield}.
Innovations for the spray simulation of the pintle injector are needed to address this issue.

Contrarily, machine learning approach, especially the \emph{Neural Networks}(NN), has demonstrated its efficiency to predict the flow fields under different conditions with a single surrogate model\cite{brunton2019machine}.
The previous researches on flow field prediction using NN are mainly focused on the data-driven method. 
Besides the indirect way using closure model\cite{ling2016reynolds,parish2016paradigm}, the field solution can also be directly obtained from the network model which is trained with a large number of samples\cite{farimani2017deep,ma2020supervised,brunton2019machine,duraisamy2019turbulence}. 
However, some predictive results obtained by data-driven methods may still exhibit considerable errors against physics laws or operating conditions\cite{farimani2017deep,ribeiro2020deepcfd,thuerey2020deep}.
Besides, In some sparse data regime, some machine learning techniques are lacking robustness and fail to provide guarantees of convergence\cite{raissi2019physics}.
For the purpose of remedying the above mentioned shortcomings of data-driven methods, the physics-driven/informed methods are proposed recently\cite{raissi2020hidden}.
By providing physics information, NN are able to directly obtain field solutions which obey physical laws and operating conditions.
In these work, \emph{Partial Differential Equations} (PDEs) was employed in the loss function to explicitly constrain the network training\cite{lu2019deepxde,ma2020combined}.

In the state-of-the-art neural networks methods, \emph{Generative Adversarial Networks} (GANs) proposed by Goodfellow et al.\cite{goodfellow2014generative}, are efficient to generate the instantaneous flow fields\cite{kim2020deep,wu2020enforcing}.
In spite of the impressive performance for unsupervised learning tasks, the quality of generated solutions by GANs is still limited for some realistic tasks\cite{arjovsky2017wasserstein}.
Besides, as shown in the training results later, the transient nature of the spray injection and liquid sheet break results in the extreme difficulty of usual networks to qualify the place and intensity of dominating characterizations.

In this paper, based on one specific type of GAN and the idea of physics constraint, a novel \emph{Generative Adversarial Networks with Physical Evaluators} (GAN-PE) framework is proposed. 
By introduction of mass conversation and spray angle model as the two evaluators, this framework has a better training convergence and predictive accuracy.
The trained model is able to simulate the macroscopic morphology and characterization of the instantaneous flow fields under different conditions. 
This paper is organized as follows. 
We firstly introduce the experimental settings and data set acquisition.
Secondly, the architecture of GAN-PE and the detailed parts are described. 
Then, the learning results of numerical experiments are presented for validation.
Finally, conclusion is drawn. 

\section{Data Set from Experiments}\label{sec:Experiments}
Our training data are extracted from the spray experimental results of the pintle injectors.

\subsection{Experimental facilities}
\begin{figure}[!htbp]
	\centering
	\subfloat[Schematic of the experimental facilities]{
		\includegraphics[height=5cm]{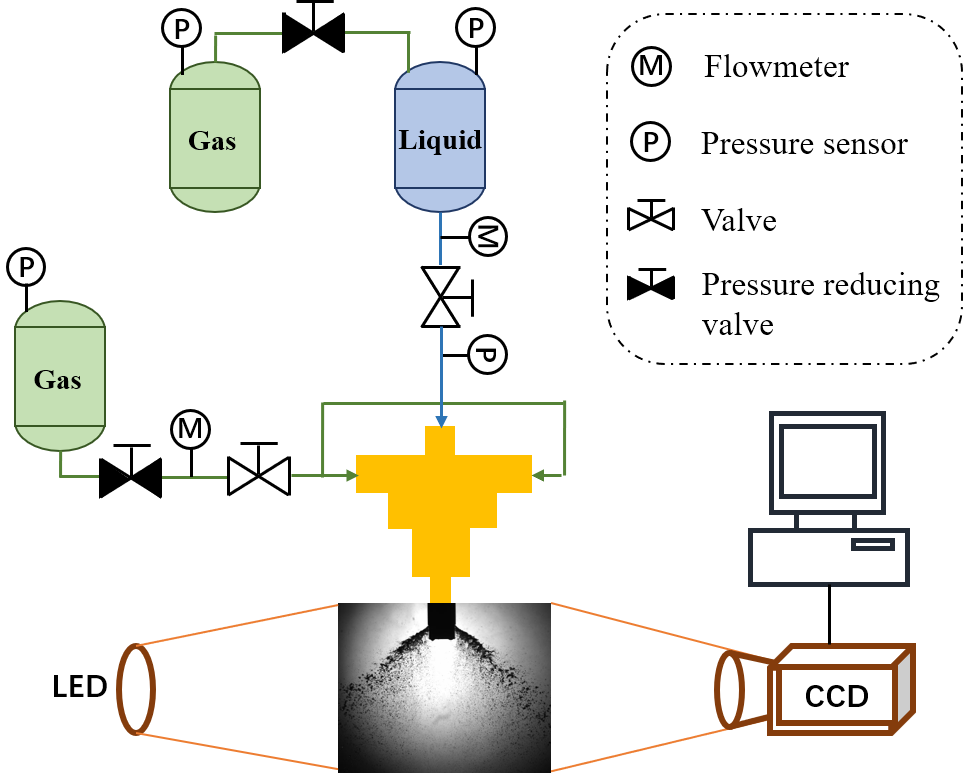}}
	\subfloat[Pintle injector]{
		\includegraphics[height=5cm]{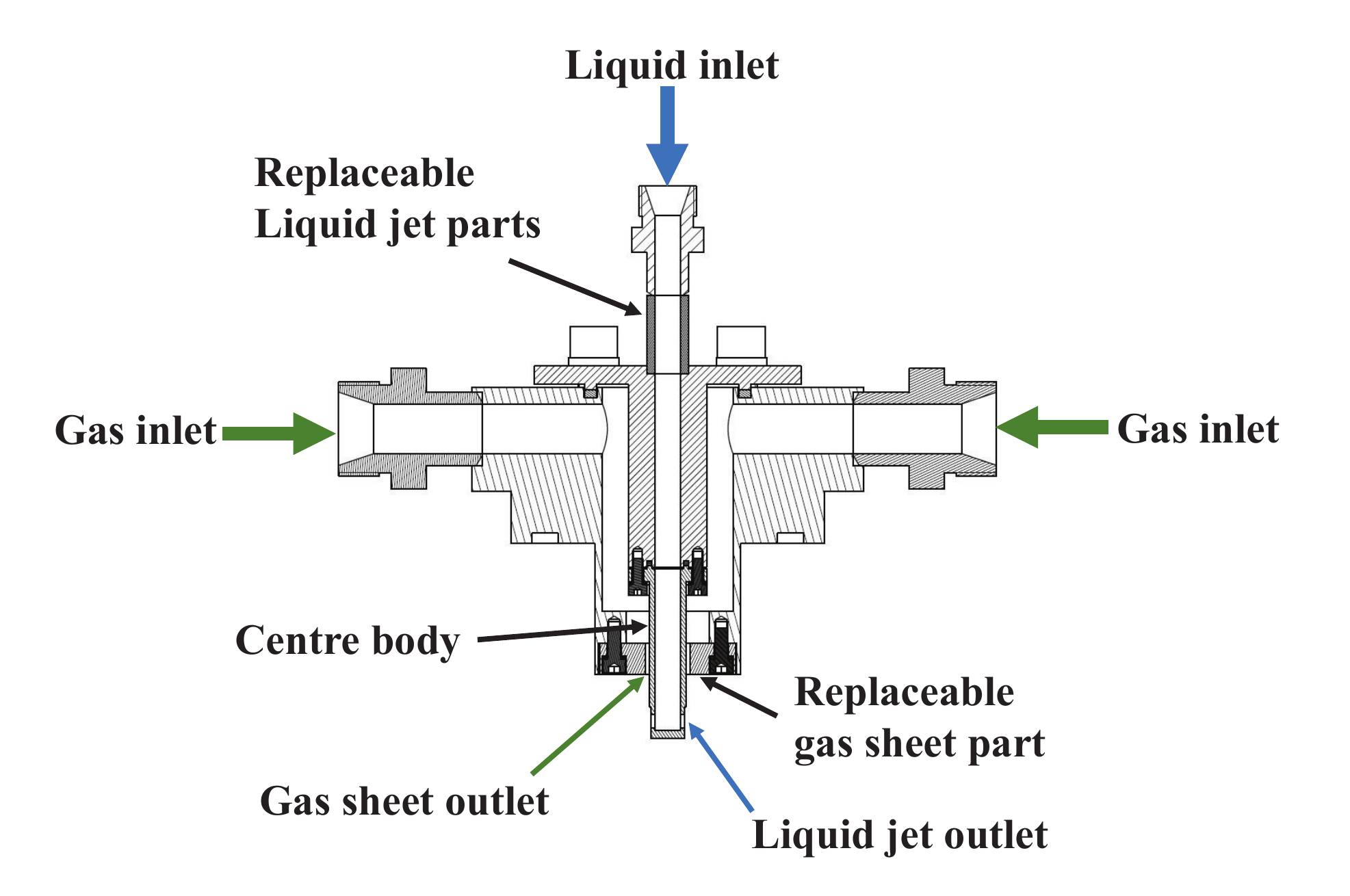}}
	\caption{Experimental facilities and devices.
		(a)Schematic of the experimental facilities. The test bench is composed of gas-liquid pintle injector, propellants feed system and control system. Spray visualization system includes a LED lamp and a high speed camera.
	    (b)Gas-liquid pintle injector within manifold.
	    The experimental injector consists of a gas manifold, a replaceable liquid manifold, a replaceable axial gas sheet adjustment annular, a central cylinder and a sleeve.
    	In order to facilitate the optical observation about the spray angle, two symmetrical radial liquid jet orifices are designed on the central cylinder.}
	\label{figs:211-Experiments}
\end{figure}

The non-reactive cold experiments were conducted at atmospheric pressure. The dry air is used as the stimulant for axial flows and filtered water as the stimulant for radial flows. 
The schematic of experimental facilities are shown in Figure \ref{figs:211-Experiments}a.  
A back-lighting photography technique is used for instantaneous spray image visualization. 
The image acquisition system consists of an LED light source, a high-speed camera and a computer. 
The exposure time is 10 \textmu s and frame rate is 50k fps.

The detailed gas-liquid pintle injector is featured in Figure \ref{figs:211-Experiments}b.
In order to study the influence of the momentum ratio on the spray angle, the experimental device is designed to use the replaceable parts.
In the experiment, the height of the radial liquid jet outlet and the thickness of the axial gas sheet are adjusted by changing the height of the sleeve and the axial gap distance, respectively.
When the liquid propellant is injected radially from the two sides of the pintle end through the manifold, the liquid columns is formed. 
These columns are broken by the axially gas propellant injection from the gap cling to the pintle. 
Finally, due to an impingement and collision, the liquid columns break and form a plane conical spray like a hollow-cone atomizer.
This design induces vigorous mixing of the gaseous and liquid propellants which yields a high combustion efficiency\cite{sakaki2017combustion}

\subsection{Data set acquisition}

The spray experiments are carried out with the throttling level $L_{t}$ of $40\%\sim80\%$.
$L_{t}$ is varied by the linear adjustment of the height of radial liquid jet outlet and the thickness of axial gas sheet. 
The radial liquid jet outlet height at throttling levels of 40\%, 60\%, and 80\% are 2mm, 3mm and 4mm, respectively.
When $L_{t}$ is fixed, the height of the radial liquid jet outlet is fixed and equal to the thickness of the axial gas sheet.
Table \ref{tab:ExpConditions} shows the operating conditions of the experimental campaign and the corresponding key specifications of the pintle injector.

\begin{table}[!htbp]
	\centering	
	\caption{Experimental operating conditions. 
		$L_{\rm open}$ and $T_{\rm gs}$ are injector opening distance and gas sheet thickness respectively. 
		$m_{\rm g}$ and $m_{\rm l}$ are the mass flow rate of gaseous and liquid propellant respectively.
		$C_{\rm TMR}$ is the momentum ratio of the two propellants.
	}
	\label{tab:ExpConditions}    
	\begin{tabular}{cccccc}
		\hline\noalign{\smallskip}
		 $L_{\rm t}$ & $L_{\rm open}(\rm mm)$ & $T_{\rm gs}(\rm mm)$ & $m_{\rm g}(\rm g/s)$& $m_{\rm l}(\rm g/s)$&  $C_{\rm TMR}$  \\
		\noalign{\smallskip}\hline\noalign{\smallskip}
		80\% & 4.0 & 4.0 & 22.17 & $18.55\sim40.54$ & $1.01\sim4.92$ \\
		60\% & 3.0 & 3.0 & 15.7  & $14.85\sim30.46$ & $1.22\sim5.14$\\
		40\% & 2.0 & 2.0 & 29.85 & \ \ $8.85\sim20.45$  & $0.98\sim5.21$\\
		\noalign{\smallskip}\hline
	\end{tabular}
\end{table}

\begin{figure}[!htbp]
	\centering
	\subfloat[Raw image]{
		\includegraphics[width=6cm,height=4.5cm]{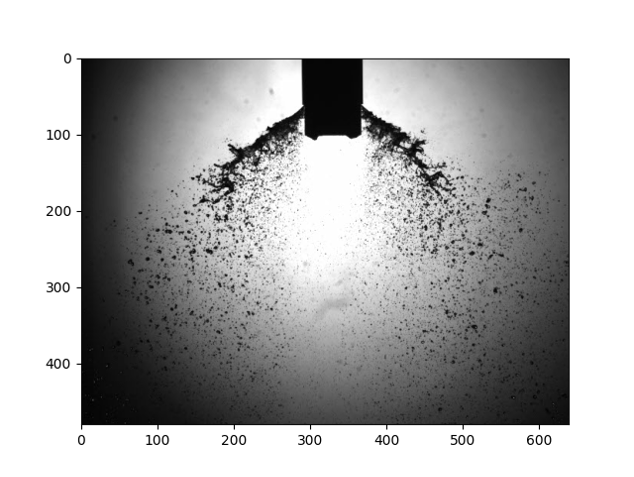}}
	\subfloat[Average]{
		\includegraphics[width=6cm,height=4.5cm]{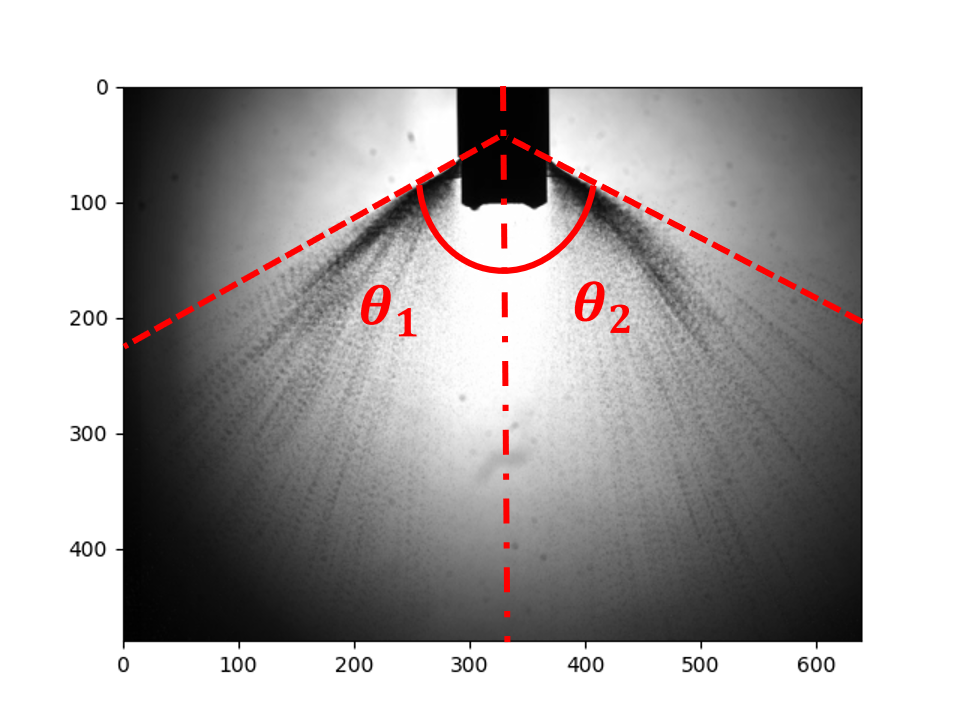}}
	\caption{Data acquisition. The resolution of the images is $640\times480$.}
	\label{figs:231-Dataacquisition}
\end{figure}

As shown in Figure \ref{figs:231-Dataacquisition}, to measure the spray angle, the spray images obtained in the experiment are post-processed to clarify the spray boundary. 
The average of 10 images is used to measure the spray angle manually.
Using this setup, the time-averaged images of sprays are obtained and the spray angles, defined as $\theta = \frac{1}{2} (\theta_{1} +\theta_{2})$, are manually-measured.
Then, the average images and the corresponding spray angles are used to train the spray angle estimator later.

The resolutions of the instantaneous spray image are $640\times480$. 
In order to reduce the training cost, the images are interpolated to the images with a resolution of $128\times128$. While the measured angle values, which represented the nature of the spray phenomenon, are fixed in spite of the image scaling.

\section{Methodology}\label{sec:Methodology}

\subsection{Overview}

\begin{figure}[!htbp]
	\centering
	\includegraphics[width=1\textwidth]{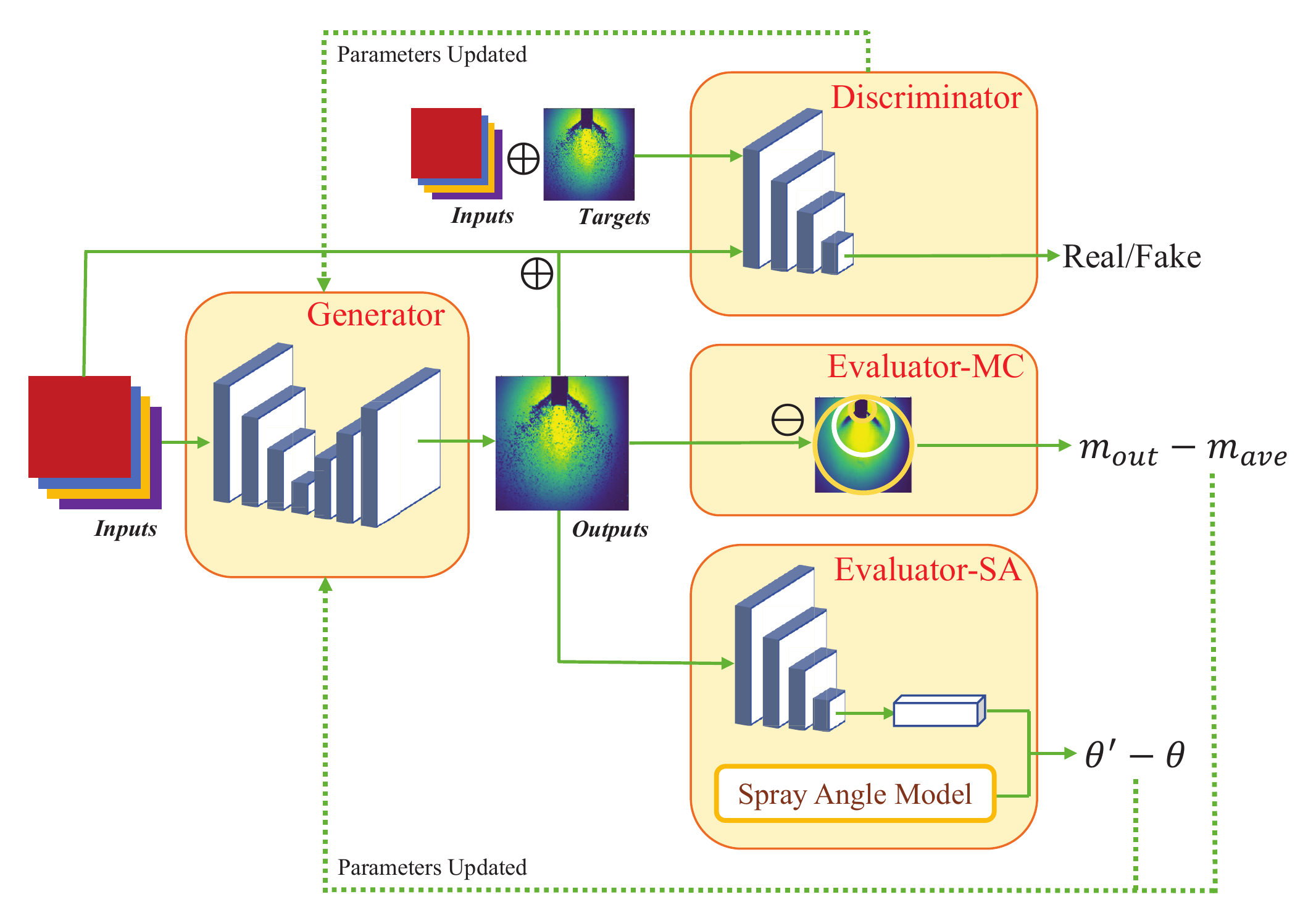}
	\caption{Schematic of the proposed networks framework. With the operating conditions, the U-net generator outputs a field solution of the spray. Then the outputs will be transferred to discriminator, mass conservation evaluator and spray angle evaluator. With the input-target pair and input-output pair, the discriminator is trained to distinguish the real and fake images. The mass conservation evaluator calculates the ring error between output and the corresponding average target. The spray angle estimator judges the angle value from the output and then compare it with theoretical one. Eventually the discriminator, mass conservation and spray angle losses are utilized to update the generator by backpropagation.} 
	\label{figs:311-Schematic}
\end{figure}

Here, a \emph{Generative Adversarial Networks with physical evaluators} (GAN-PE) framework is proposed. 
As shown in Figure \ref{figs:311-Schematic}, The GAN-PE is composed of 4 parts, \emph{Generator} ($G$), \emph{Discriminator} ($D$) and two physical evaluators.
After the initial field solution is generated by $G$, there are three parts are employed to guarantee the output being an accurate field solution. 
GANs is the base of the proposed networks framework, the $G$ captures the real spray data distribution which is corresponding to the operation conditions, and the $D$ estimates
the probability that a condition-sample pair came from the training data rather than $G$.
There are also two evaluator designed to improve the performance of GANs.
The first \emph{Mass Conservation Evaluator} ($E_{\rm MC}$) is used to improve the generation robustness and training convergence.
The second \emph{Spray Angle Evaluator} ($E_{\rm SA}$) is used to improve the predicting accuracy in the specific operating conditions.
Fed with the outputs from $G$, the losses of $D$, $E_{\rm MC}$ and ${E_{\rm SA}}$ are calculated respectively. 
After that, the backpropagation is applied to adjust the U-net CNN of $G$ to generate a new spray field that more satisfies the conditions and prier physics knowledge. 
After enough iterations, the network will be able to generate ‘correct’ spray field.

\subsection{GAN}

\noindent  
\emph{Generator}

From inputs towards outputs, the network of $G$ consists of two parts: encoding and decoding\cite{ronneberger2015u} .
In the encoding process, the operating conditions $L_{\rm open}$, $T_{\rm gs}$, $m_{\rm g}$ and $m_{\rm l}$ are resized as four matrices for progressively convolutional down-sampling with corresponding kernels. 
By this way, the matrices with a size of $128 \times 128$ is reduced into one liner data pool consisting of 512 neurons.
Then the decoding part works in an opposite way, which can be regarded as an inverse convolutional process mirroring the behavior of the encoding part. 
Along with the increase of spatial resolution, the spray fields are reconstructed basing on the data pool by up-sampling operations.
In addition, there are the concatenation of the feature channels between encoding and decoding.
More details of the U-net architecture and convolutional block, including active function, pooling and dropout, are referred to Ref. \cite{ma2020combined}. 

The weighted loss function considering the following discriminator and evaluators is written as
\begin{equation}
\label{con:Loss}
\mathcal{L}(D, E_{\rm MC}, E_{\rm SA})=
\mathcal{L}_{D}
+ {\rm \alpha}\mathcal{L}_{E_{\rm MC}} 
+ {\rm \beta}\mathcal{L}_{E_{\rm SA}},
\end{equation}
where $\mathcal{L}_{D}$, $\mathcal{L}_{E_{\rm MC}}$ and $\mathcal{L}_{E_{\rm SA}}$ are the loss terms that calculated by $D$, $E_{\rm MC}$ and $E_{\rm SA}$ respectively. 
Also, ${\rm \alpha}$ and ${\rm \beta}$ are the constant hyperparameters which are tuned to adapt the scales of these loss terms. 
After proper training, the generator is able to map a spray sample from a random uniform distribution to the desired distribution which obey the physical knowledge and conditions.

\noindent  
\emph{Discriminator}

Then $D$ is used to feed the possibility of that samples come from the training rather than generation distribution back to $G$.
We use \emph{Least Squares Generative Adversarial Networks} (LSGANs) settings to train the $D$ and the $G$ simultaneously\cite{mao2017least}. 
This special type of GANs helps to remedy the gradients vanishing by using the least square loss function instead of the sigmoid cross entropy loss function\cite{karras2017progressive}.

Here, $D$ is modified by the encoder of the $G$, which means the generations are dawn-sampling by the re-convolutional calculation so that the spray field information are concluded into the linear 1-D tensor.
And then this 1-D data pool will be used to be trained to maximize the probability of assigning the correct label (real/fake) to both training targets and generating solutions.
Similar with the work in Ref.\cite{mirza2014conditional}, we use the input-output pair to feed $D$ instead of only $G$'s outputs in the random image generation tasks. 
The operating conditions and the outputs are concatenated as the different feature channels in a unique 4-D data tensor.
By this way, the $D$ helps to judge whether the outputs accord with the corresponding conditions, not only having right spray morphology.
  
The loss functions for LSGANs are defined as
\begin{equation}
\label{con:LossGAN}
\begin{split}
\min_{D}V_{\rm GAN}(D) 
&
=\frac{1}{2}\mathbb{E}_{\boldsymbol{x}\sim p_{data}(\boldsymbol{x})}[(D(\boldsymbol{x})-b)^2] 
+\frac{1}{2}\mathbb{E}_{\boldsymbol{z}\sim p_{\boldsymbol{z}}(\boldsymbol{z})}[(D(G(\boldsymbol{z}))-a)^2]
\\
\min_{G}V_{\rm GAN}(G) 
&
=\frac{1}{2}\mathbb{E}_{\boldsymbol{z}\sim p_{\boldsymbol{z}}(\boldsymbol{z})}[(D(G(\boldsymbol{z}))-c)^2],
\end{split}
\end{equation}
where $\boldsymbol{x}$ is the training data and $\boldsymbol{z}$ is the input variables.
Also, $a$ and $b$ are the labels for fake data and real data respectively, and $c$ denotes the possibility that $G$ wants $D$ to believe for fake data. Here, we apply $a=0$ and $b=c=1$. 
So, $\mathcal{L}_{D}$ is equal to the second part of Equation (\ref{con:LossGAN}).

\subsection{Evaluators}

\noindent  
\emph{Mass conservation evaluator}

\begin{figure}[!h]
	\centering
	\subfloat[Generated images]{
		\includegraphics[width=0.4\textwidth, angle=0]{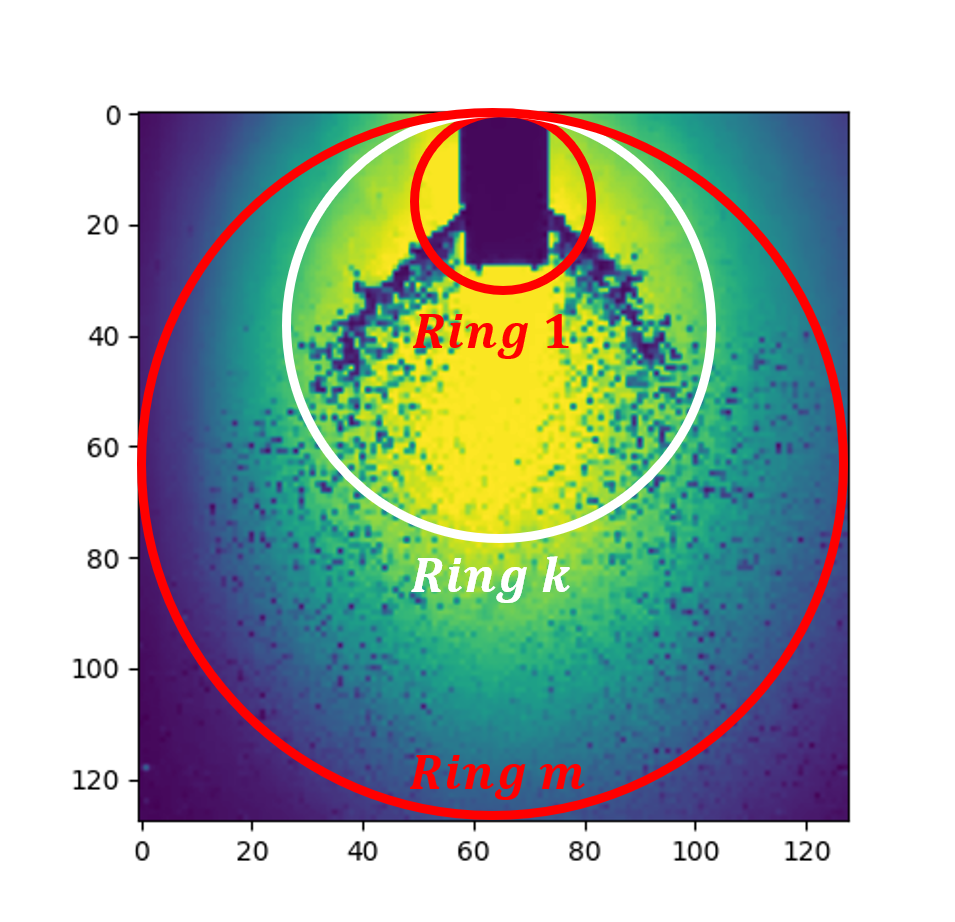}}
	\subfloat[Average target]{
		\includegraphics[width=0.4\textwidth, angle=0]{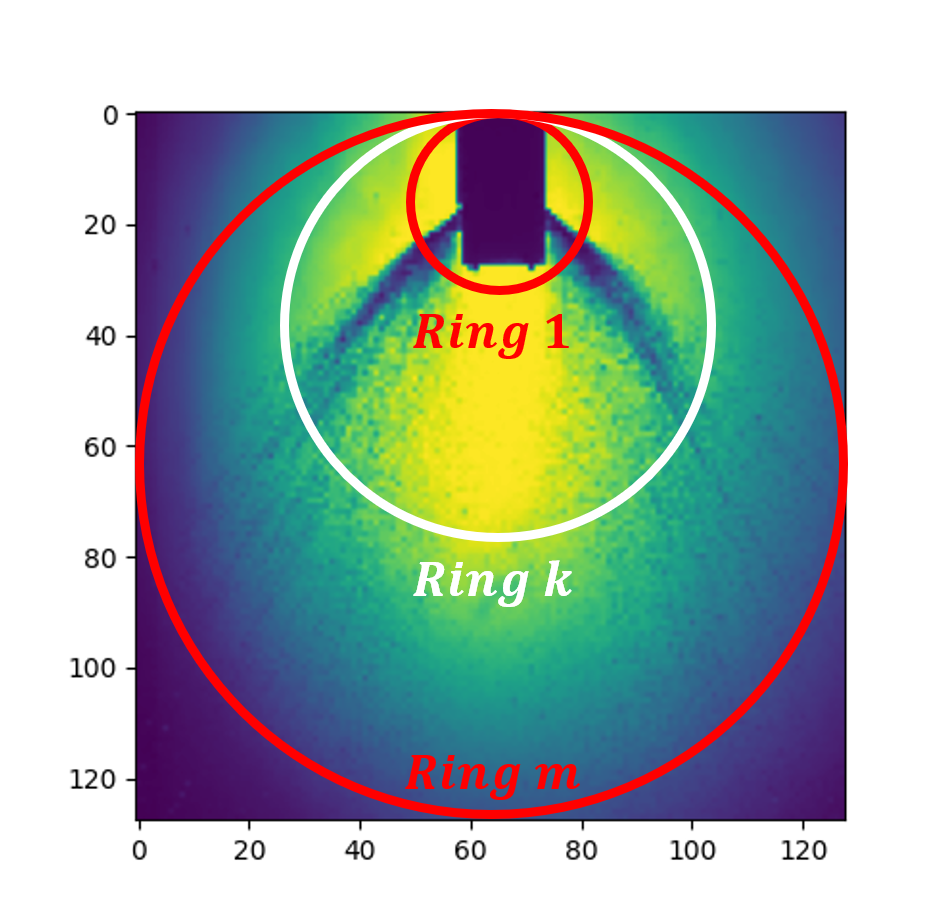}}
	\caption{Mass Conservation. The resolution of the images is $128\times128$.}
	\label{figs:331-MassConservation}
\end{figure}

As shown in Figure \ref{figs:331-MassConservation}, we assume that there are a few rings with different diameters that are tangent at the middle point of the upper boundary in both generating images and the average images. 
Following the definition of ``L1loss'' which is widely used in machine learning community, we define a mass conservation loss here. 
The difference is that the former measures the sum of absolute error between each element in the generation and target\cite{paszke2019pytorch}, but ours is calculating firstly a gray value summation of every element in one concerning ring and then the summation of absolute error between the corresponding rings in the generation and target.
The idea is from that the spray phenomenon obeys mass conservation law, i.e. the mass fluxes of droplets through one specific ring in every instantaneous frame are equivalent.
Here, the ``mass flux'' is a broader concept which also involves the background shadow.

The mass conservation error, i.e. the loss term from $E_{\rm MC}$, is defined as
\begin{equation}
\label{con:LossE_MC}
\mathcal{L}_{E_{\rm MC}}=
\begin{dcases}
\sum_{k=1}^{m} { {\left|\sum_{i=1}^{n} {x}_i - \sum_{j=1}^{n} {y}_j \right|}_k  }
&
\mathcal{E}\geq\mathcal{E}_{\rm thr}
\\
0 
&
\mathcal{E}<\mathcal{E}_{\rm thr}
\end{dcases},
\end{equation}
where $x$ and $y$ are the gray values in generated images and average targets respectively. Also, $m$ is the amount of the concerning rings and $n$ is the amount of data points in one concerning  rings in the matrix.
$\mathcal{E}$ is the error between the output and average matrix and defined as
\begin{equation}
\mathcal{E}=
\sum_{i=1}^{N}\left|{x}_i - {y}_i \right|,
\end{equation}
where $N$ is the amount of all the data points in the matrix.
To improve the generation randomness and in view of the error caused by light transmission and reflection, loss threshold $\mathcal{E}_{\rm thr}$ is introduced herein. Once the loss is less than the threshold, this loss term will be ignored.   

\noindent  
\emph{Spray angle evaluator} 	
\label{sec:33a-SprayAngle}

This evaluator is composed of two parts, one is the theoretical model of spray angle, the another is a CNN encoder to estimate the spray angles from the generated field solutions.

\begin{figure}[!htbp]
	\centering
	\includegraphics[width=0.5\textwidth]{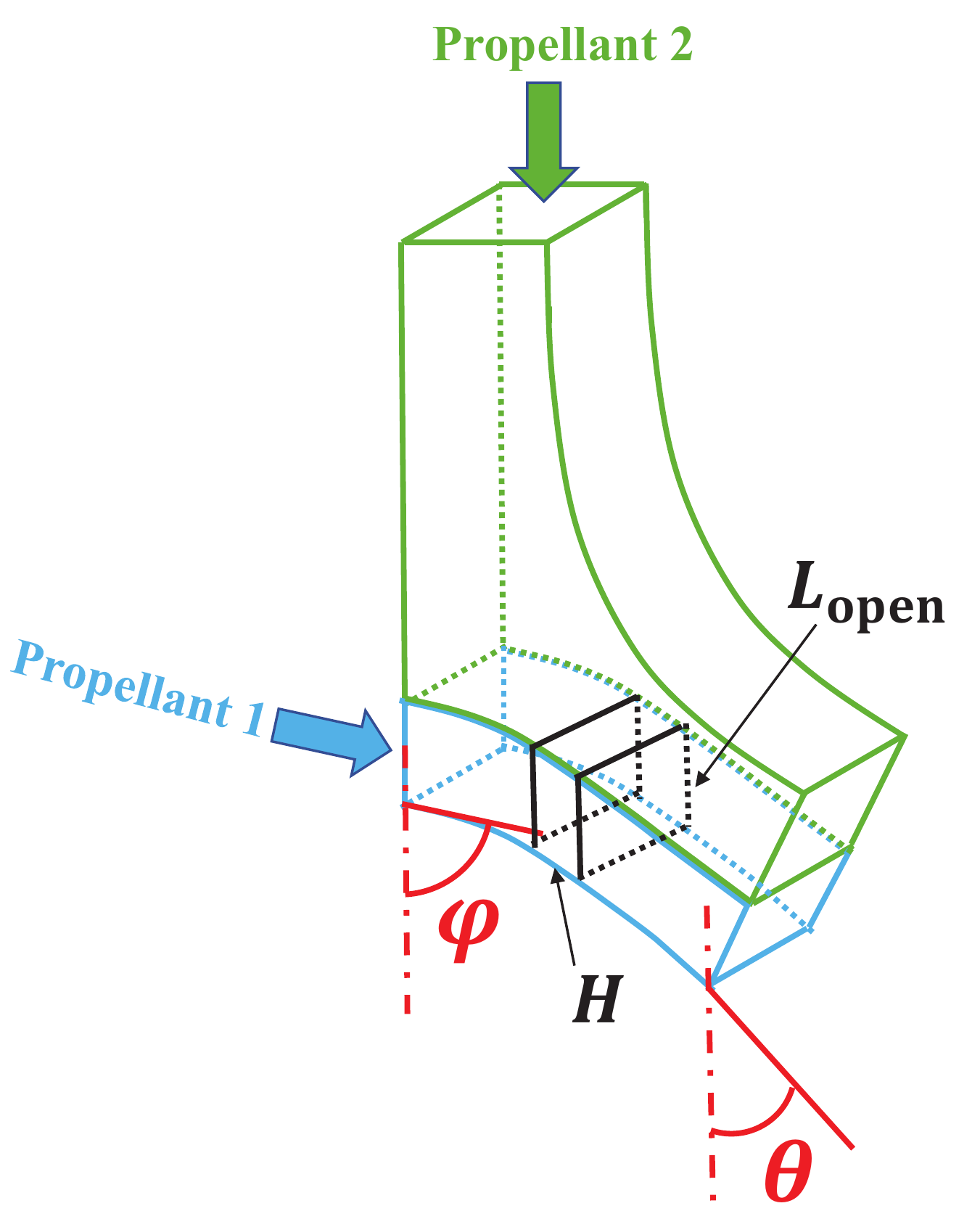}
	\caption{Spray angle model}
	\label{figs:332-SprayAngleModel}
\end{figure}

The schematic diagram of the theoretical model of spray angle is shown in Figure \ref{figs:332-SprayAngleModel}. 
Several basic hypotheses must be declared before carrying out the theoretical analysis.
(a). An element of fluid emerging from the jet exit is assumed to have the constant length and width equal to the jet exit length $L$ and width $W$, as it moves along the trajectory.
(b). Liquid jet deformation, evaporation and droplet dispersion are ignored.
(c). As the fluid element exits the slot, it has an initial velocity and an initial angle equal to the central propellant deflection angle $\varphi$.
(d). The aerodynamic force is assumed to be parallel to the gas sheet flow direction, and the movement direction of the fluid element is deflected.
(e). The spray angle is assumed to be equal to the slope of the liquid jet at the thickness point of the gas film.
(f). Surface tension, gravity, friction, heat transfer and phase change are ignored.

The axial momentum equation can be written as follows:
\begin{equation}\label{con:1}
\frac{1}{2} \rho_{\rm g}\left(u_{\rm g}-u_{\rm l} \rm{cos\varphi}\right)^{2} \emph{W} \emph{H}
= \rho_{\rm l} \emph{W}\emph{L}_{\rm open} \emph{H} \frac{d\emph{u}}{d\emph{t}},
\end{equation}
so that integration of the axial momentum equation with respect to time is
\begin{equation}\label{con:2}
u_{\rm l}
=
\frac{1}{2} \frac{\rho_{\rm g}}{\rho_{\rm l}} \frac{\left(u_{\rm g}-u_{\rm l}  \rm cos\varphi\right)^{2}}{\emph{L}_{\rm open}}t
+u_{\rm l}\rm cos\varphi,
\end{equation}
with $u_{\rm l}=\frac{\rm d\emph{u}}{\rm d\emph{t}}$, a second integration with respect to time is
\begin{equation}\label{con:3}
x
=
\frac{1}{4}\frac{\rho_{\rm g}}{\rho_{\rm l}} \frac{\left(u_{\rm g}-u_{\rm l}  \rm cos\varphi\right)^{2}}{\emph{L}_{\rm open}} \left(\frac{y}{u_{\rm l} \rm sin \varphi}\right)^{2}
+ \frac{ y u_{\rm l} \rm cos \varphi }{ u_{\rm l} \rm sin \varphi}.
\end{equation}
For the collision between the gas sheet and the rectangular liquid jet, the momentum ratio is
\begin{equation}\label{con:4}
C_{\rm TMR}
=
\frac{\dot{m_{\rm l}} v_{\rm l}}{\dot{m_{\rm g}} v_{\rm g}}
=
\frac{\rho_{\rm l} v_{\rm l}^{2} A_{\rm l}}{\rho_{\rm g} v_{\rm g}^{2} A_{\rm g}}
=
\frac{\rho_{\rm l} v_{\rm l}^{2} WL_{\rm open}}{\rho_{\rm g} v_{\rm g}^{2} \emph{W} \emph{H}}
=
\frac{\rho_{\rm l} v_{\rm l}^{2} L_{\rm open}}{\rho_{\rm g} v_{\rm g}^{2} H}.
\end{equation}
Equation \ref{con:3} could then be expressed in terms of the momentum ratio $C_{\rm TMR}$ as follows
\begin{equation}\label{con:5}
x
=
\frac{1}{4C_{\rm TMR}H\rm{sin}^{2}\theta}
\left(1-\frac{u_{\rm l}  \rm cos\theta}{u_{\rm g}}\right)^{2} y^{2}
+
\frac{\rm cos \theta }{\rm sin \theta }y,
\end{equation}
where the slope of the liquid jet $\theta$ at the thickness of the gas sheet could be written as
\begin{equation}\label{con:6}
\theta
=
90^{\circ} \frac{1}{2C_{\rm TMR}\rm{sin}^{2}\varphi} 
\left(1-\frac{u_{\rm l}  \rm cos\varphi}{u_{\rm g}}\right)^{2}
+
\frac{\rm cos \varphi }{\rm sin \varphi }.
\end{equation}

The theoretical model assumes that the liquid jet does not deform, but in reality it will deform under aerodynamic forces, which results in a reduction of the effective momentum of the liquid jet. 
Therefore, the liquid jet deformation factor $\gamma$, which is obtained through the experimental results, is introduced to modify the spray angle theoretical model. 
The Eq(\ref{con:6}) is rewritten as
\begin{equation}\label{con:7}
\theta
=
\gamma \left(90^{\circ} \frac{1}{2C_{\rm TMR}\rm{sin}^{2}\varphi} 
\left(1-\frac{u_{\rm l}  \rm cos\varphi}{u_{\rm g}}\right)^{2}
+
\frac{\rm cos \varphi }{\rm sin \varphi }\right).
\end{equation}

In the field of medical image analysis, the machine learning approach, especially the deep neural networks, has been employed for automated scoliosis assessment\cite{yang2019development}. In these work, the X-ray images are fed into the neural networks estimator and the spinal Cobb angles are obtained\cite{cai2020computational,wu2018automated}.     
Similarly, inside the $E_{\rm SA}$, there is a well-trained spray angle estimator to output the angle values from the predictive images.
The architecture of this CNN encoder is like the $D$, except added one liner layer in the end to output the estimated spray angle $\theta^{\prime}$.

The loss term from $E_{\rm SA}$ is calculated as
\begin{equation}\label{con:LossE_SA}
\mathcal{L}_{E_{\rm SA}}
=
\theta^{\prime}
-
\theta
.
\end{equation}

\section{Results}\label{sec:Results}

\subsection{Model validation} 
\begin{figure}[!htbp]
	\centering
	\subfloat[Experiment]{
		\includegraphics[width=0.3\textwidth
		]{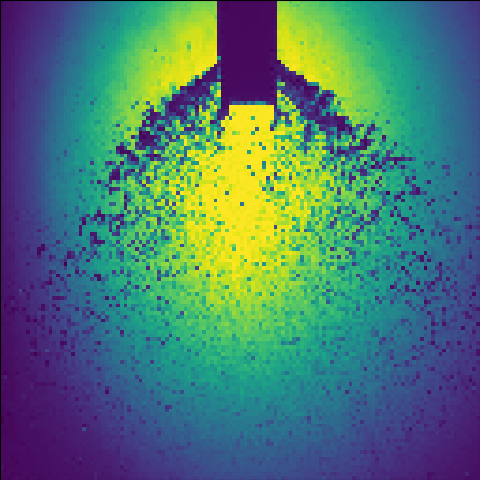}}
	\subfloat[CNN]{
		\includegraphics[width=0.3\textwidth]{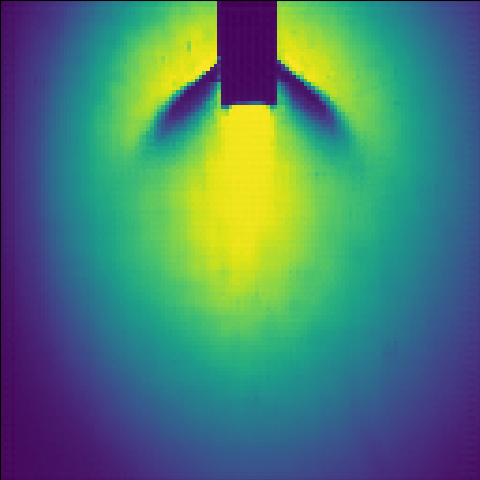}}
	\subfloat[Original GAN]{
		\includegraphics[width=0.3\textwidth]{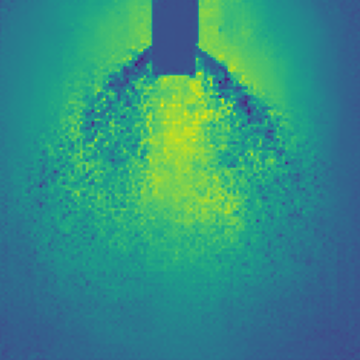}}
	\caption{Comparison of CNN generator and original GAN. $ L_{\rm open}= T_{\rm gs} = 4 \rm mm $, $m_{\rm l} = 35.81 \rm g/s$, $m_{\rm g} = 22.17 \rm g/s $ .}
	\label{figs:411-CNNvsGAN}
\end{figure}
Figure \ref{figs:411-CNNvsGAN} shows the generated results in one typical operating condition of CNN generator and original GAN.
The L1 loss used by CNN generator compares the difference between the generations and targets. 
This absolute error loss performs very well in some steady or mean state field prediction tasks, such as the work in Ref. \cite{thuerey2020deep, ma2020combined}. 
However, when the training cases have  a multi-modal distribution, this loss will fail down.
In our spray field prediction task, although the morphology under one specific operation condition are similar, but the detailed droplet distributions are distinguishable.
So the spray field solution actually has many possibilities and the prediction in every iterative step should be one of them.
But the L1 loss averages all the possibilities and produces a very blurry average image instead.
However, the discriminator in GANs which can be regarded as the loss of generator is not an explicit loss function.
Instead of the pixel-wise loss, $D$ is an approximation loss and it denotes the overall spray morphology which discriminates between the real and fake data distributions. 


In the training process of GANs, the generator and discriminator have to been balanced trained and the convergence is often an unstable state.
For the spray simulation task, the discriminator have difficulties to capture the detailed feature of the small droplets.

\begin{figure}[!htbp]
	\centering
	\subfloat[Experiment]{
		\includegraphics[height=0.3\textwidth]{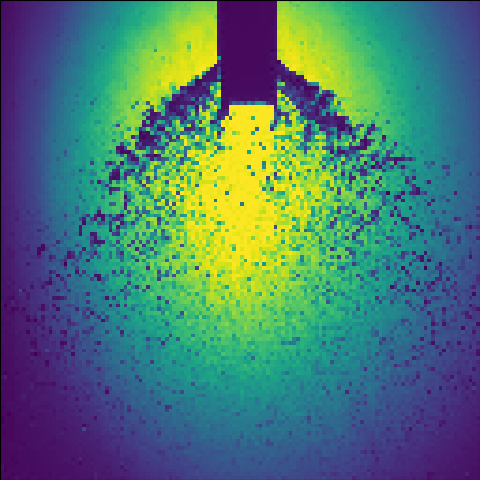}}
	\subfloat[GAN with $E_{\rm MC}$]{
		\includegraphics[height=0.3\textwidth]{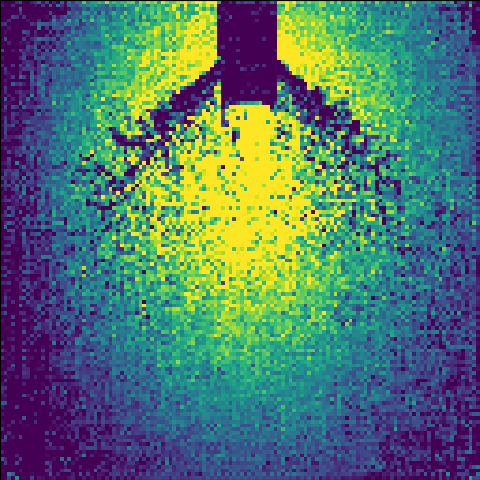}}
	\\
	\subfloat[Original GAN]{
		\includegraphics[height=0.3\textwidth]{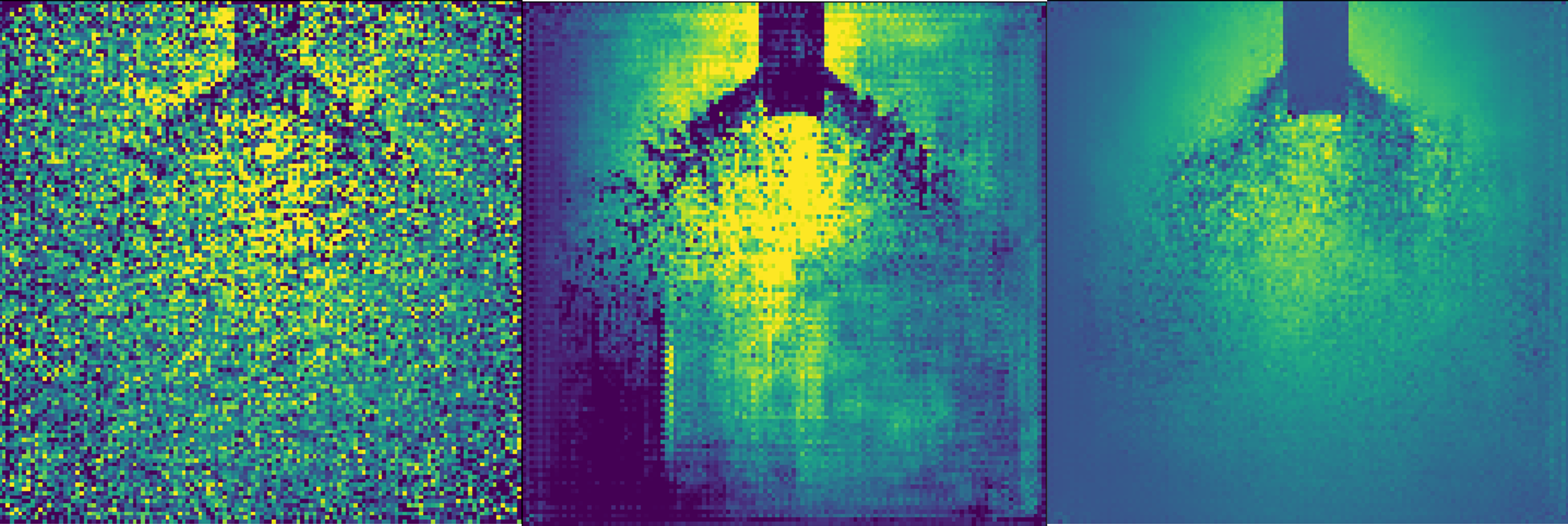}}

	\caption{The comparison between the results of with or without $E_{\rm MC}$.}
	\label{figs:421-withoutMCE}
\end{figure}

The $\mathcal{L}_{D}$ has a possibility of becoming less meaningful through the training process.
And the $G$ will update itself based on the random feedback and the quality of generation may collapse.
The $G$ outputs low-quality images through many epochs, some of them shows faint spray pattern in the background but are easily identified as fake.
It will be very easy for the discriminator to distinguish the targets and generation so the values of the loss from $D$ drop to zero rapidly. 
The comparison of generated spray images from different framework demonstrates the superior performance of the mass conservation evaluator as shown in Figure \ref{figs:421-withoutMCE}.
In some generated images, the background is not agree with the real target, the introduction of the $\mathcal{L}_{E_{\rm MC}}$ helps $G$ identifying the position and intensity of the droplet as well as the background shadow.


In our framework, when getting the predicted spray field, all the output matrices in one batch will be averaged to one image.
Then this average image will be used to feed the down-sampling CNN to obtain the spray angle of this batch.
Here, we use the average images from experiments as test samples for the validation of the spray angle estimator.
Table \ref{Tab03} shows the comparison of spray angles obtained by \emph{Manual Measurements} (MM) and CNN estimator.
With the increase of $C_{\rm TMR}$ , the deviations between the two 
tend to be smaller. 
The error of all the test cases are less than 5.8\%. 

\begin{table*}[!htbp]
	
	\scriptsize
	
	\centering
	
	\caption{Spray angle estimation by manual measurement and CNN.}
	
	\label{Tab03}
	
	\begin{tabular}{cccccccccc}
		
		\toprule
		
		\multirow{2}{*}{$C_{\rm TMR}$} & \multicolumn{3}{c}{$L_{\rm t}=80\%$} & \multicolumn{3}{c}{$L_{\rm t}=60\%$} & \multicolumn{3}{c}{$L_{\rm t}=40\%$} \\
		
		\cmidrule(r){2-4} \cmidrule(r){5-7} \cmidrule(r){8-10}
		
		&  MM      &  CNN   &   Error(\%)
		
		&  MM      &  CNN   &   Error(\%)
		
		&  MM      &  CNN   &   Error(\%)  \\
		
		\midrule
		
		$0.52\sim0.56$             &31.27                          & 29.70                    & 5.0                   & 29.31           & 28.04           & 4.3          & 26.05           & 27.57           & 5.8         \\
		
		$0.79\sim0.86 $             & 36.32           & 35.10           & 3.4                  & 33.12           & 34.54           & 4.3         & 31.59           & 32.61           & 3.2         \\
		
		$0.98\sim 1.22 $             &40.67                          & 41.98                    & 3.2                  & 34.89           & 35.60           & 2.0          & 34.67           & 35.15           & 1.4        \\
		
		$1.29\sim 1.34 $             &41.98                          & 40.37                  & -3.8                 & 39.21           & 39.34          & 0.3        & 36.36      & 37.0          & 1.75      \\
		
		$1.50\sim 1.70$             &44.37                          & 45.99               & 3.7                   & 42.87           & 43.37     & 1.2          & 39.43           & 38.08          & -3.4         \\

		$1.98\sim 2.04$             &45.28                          & 47.45                    & 4.8                   & 43.43           & 44.62          & 2.7         & 41.01   & 41.47          & 1.1          \\
		
		$2.56\sim 2.69$             &49.88                          &49.43               & -0.9                   & 47.34         &45.46        & -4.0          & 44.64           & 44.39          & -0.6            \\
		
        $2.90\sim 3.22$             &50.36                         & 50.02                  & -0.7                  & 48.13           & 47.07     & -2.2       & 44.98           &45.64        & 1.5            \\

        $3.33\sim 3.39$             &51.32                         & 52.92     &3.1            &48.12                   & 48.75         & 1.3    & 46.01     & 46.64           & 1.4              \\

        $3.83\sim 4.12$             &52.18                          & 52.55                    & 0.7                   & 48.84           & 51.09          & 4.6         & 447.64           & 48.98          & 2.8           \\

        $4.50\sim 4.88$             &54.39                          & 53.92                    & -0.9                   & 49.34          & 50.57        & 2.5         & 48.80           & 49.60         & 1.7          \\

        $5.12$             &54.55                          & 55.96                & 2.6                   & 49.41           & 50.75         & 2.7        & --           & --          & --            \\			
		
		\bottomrule
		
	\end{tabular}
	
\end{table*}

After this CNN estimator being trained, it only takes some milliseconds to output a angle value which is according to the operating conditions of this batch.
Then, the estimated value is employed in the $\mathcal{L}_{E_{\rm SA}}$ to update $G$.   
Due to the quick decrease in the beginning of the training, this error is no longer affect the $G$, only except for the generated solutions with a paradoxical angle.

\subsection{Predictions} 

Literature shown that the macroscopic morphology study is important to characterize a spray\cite{luo2018experimental}.
Here, the simulated spray morphology is analyzed and compared with the experimental results.
\begin{figure}[!htbp]
	\centering
	\subfloat[$L_{\rm t}=80\%$]{
		\includegraphics[width=12cm]{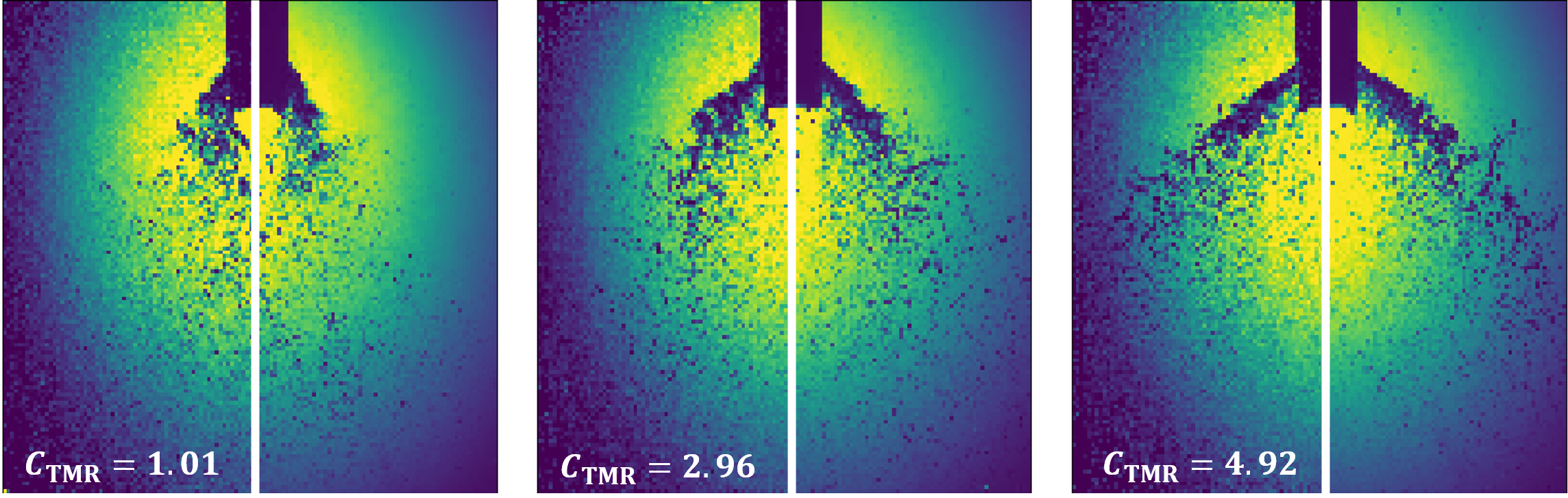}}
	\quad
	\subfloat[$L_{\rm t}=60\%$]{
		\includegraphics[width=12cm]{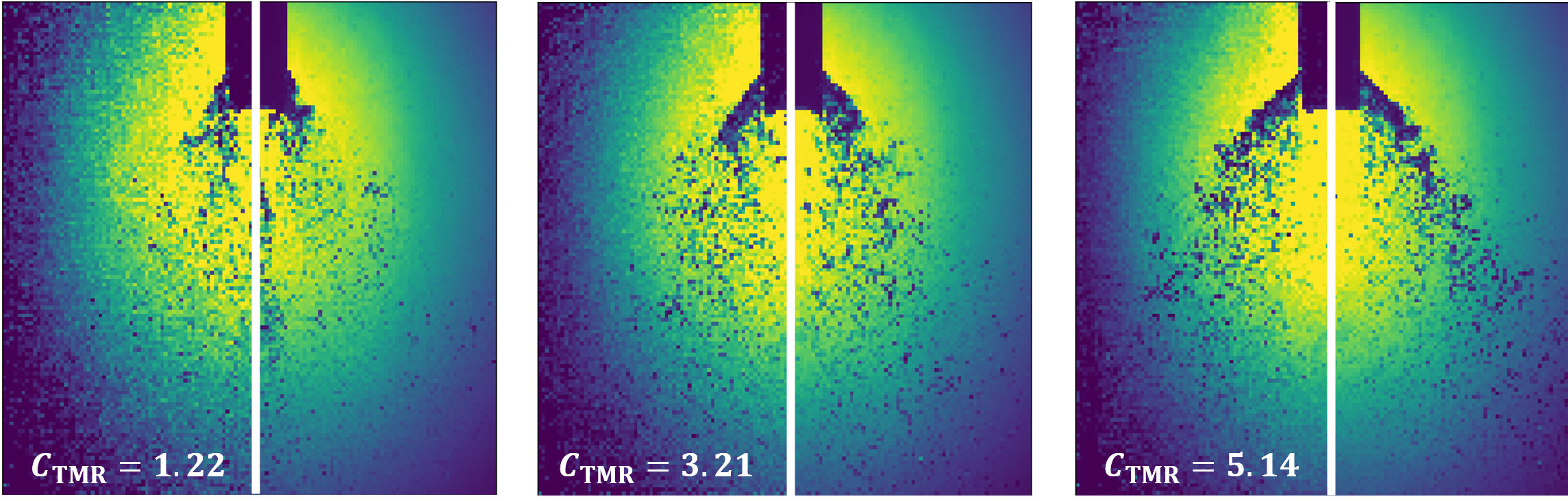}}
	\quad
	\subfloat[$L_{\rm t}=40\%$]{
		\includegraphics[width=12cm]{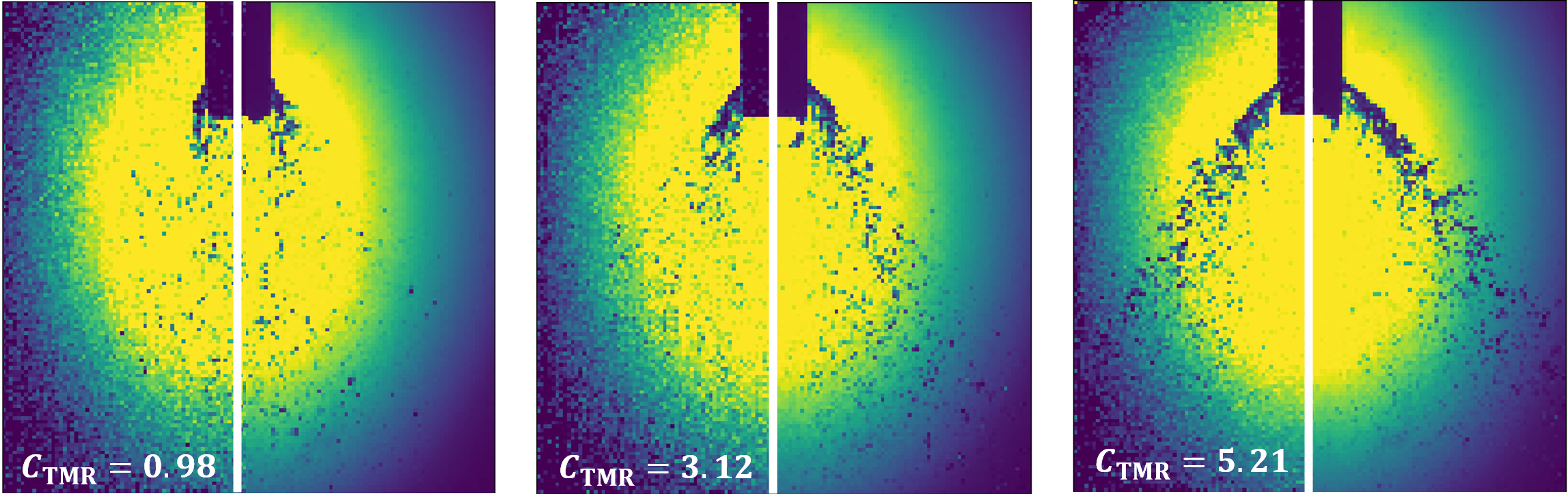}}
	
	\caption{Simulated and experimental spray morphology
		under different throttling level and momentum ratio.
		The left half part is the simulated spray obtained by GAN-PE, and the right half part is the corresponding experimental high-speed image with the same size scale.}
	\label{figs:441-Field}
\end{figure}

Figure \ref{figs:441-Field} compares the simulated and experimental spray morphology under different operating conditions.
The typical spray morphology, i.e. liquid column formation, breakup of the column, lateral expansion of the spray, can be clearly noted.
The generated spray field shows that the droplets experience a size reduction before approaching a uniform tiny size distribution because of the impingement and collision. 
As the figure shown, the generated spray is fairly comparable with the experimental results, meanwhile the hollow-cone-shaped profiles with a specific spray angle are well reproduced.
For those cases that are out of the training domain, the last column in the figure, the generation is also present a similar good quality compared with those inside.   
Imperfectly, the background still represents grainy and the values of adjacent data points are not as continuous as those in the real images.
However, all in all, the simulation succeeds to report the macroscopical morphology of the spray.

\begin{figure}[!htbp]
	\centering
	\includegraphics[width=1\textwidth]{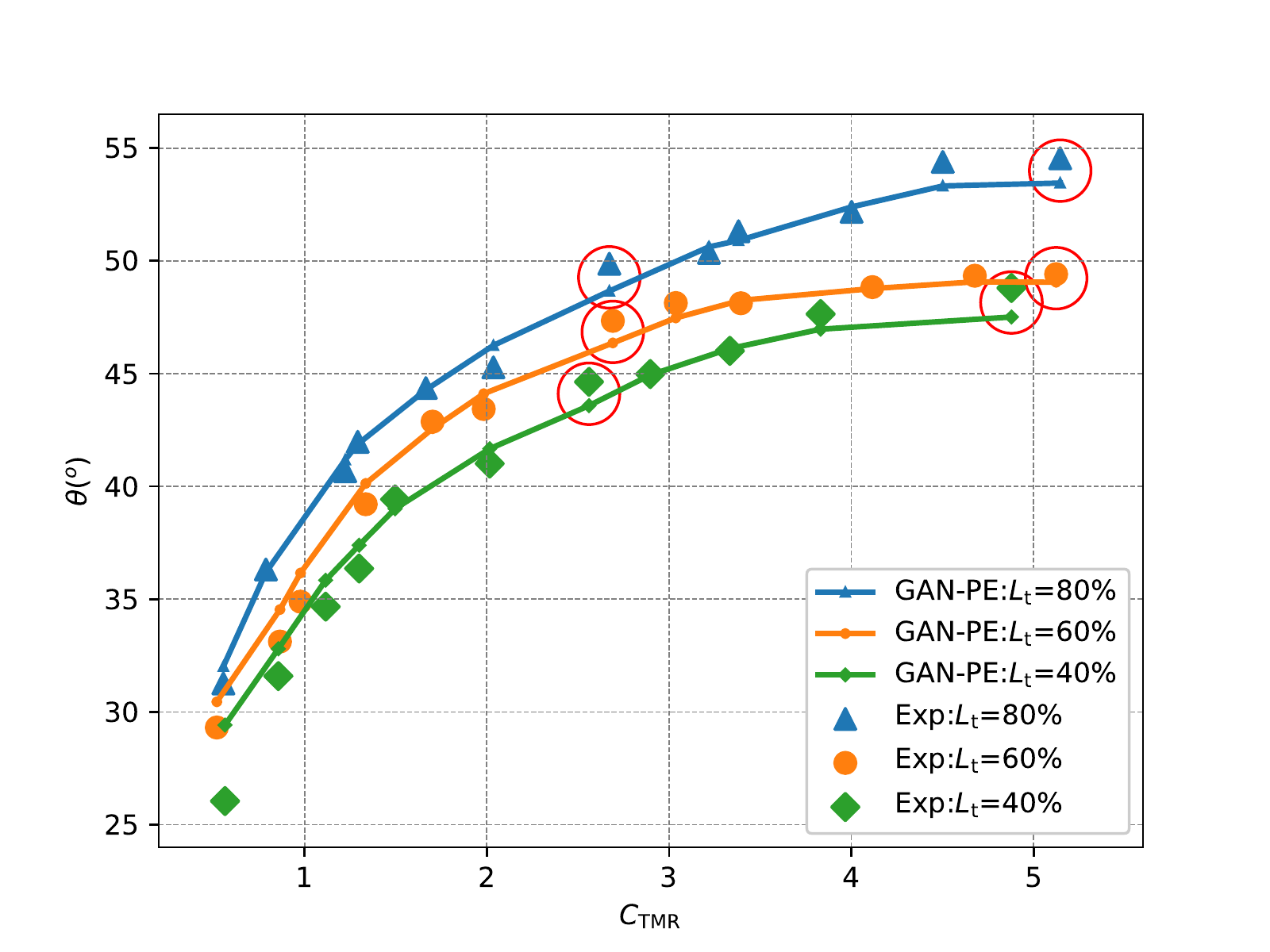}
	\caption{The comparison of spray angle between GAN-PE and experiments. The red circles show the cases which are out of learning domain.}
	\label{figs:442-SprayAngleCurves}
\end{figure}

Figure \ref{figs:442-SprayAngleCurves} shows the curves of the spray angle versus momentum ratio at different throttling levels, it can be observed that the GAN-PE results coincide with the experimental results well in a wide momentum ration range.
According to theoretical analysis which has been explained in Section \ref{sec:33a-SprayAngle}, the spray angle is mainly determined by the momentum ratio, and the simulated solutions also represent it.
For the cases which under a small momentum ratio, due to the difficulties to estimate the small spray angle, the predicted ones have a relatively bigger errors and the values of different throttling levels are close to each other. 
For the test cases which are not in the learning domain, the results have a small deviation and all the predicted angle values are less than the manually-measured ones.
It is because these test cases are not constrained by the targets so the prediction have a trend to approach the mean value of the adjacent operating points which is happened to be larger.

\section{Conclusion}\label{sec:Conclusion}
In this paper, we proposed a novel deep learning framework constrained by physical evaluators to directly predict spray solutions based on generative adversarial networks.
The normal discriminator and the mass conservation and spray angle evaluators are used to constrained the CNN to generate the spray solution, including macroscopical morphology and spray angle.
The former evaluator is able to improve the training convergence and the latter one helps obtaining more accurate solutions that are consist with the operating conditions.
It is noteworthy that the related network architecture and spray problem are generic and the proposed framework is potentially suitable for other fluid field simulations which have proper prior physics knowledge. 
Further research will be carried out for spray droplet size analysis  and prediction with the present network framework.

\section{Acknowledgement}

Hao Ma (No. 201703170250) is supported by China Scholarship Council when he conducts the work this paper represents.
Chi Zhang would like to express his gratitude to Deutsche Forschungsgemeinschaft for his sponsorship under grant number DFG HU1527/6-1 and DFG HU1527/10-1.

\clearpage

\bibliography{mybibfile4}

\end{document}